\documentclass{JAC2000}
\usepackage{epsfig}
\usepackage{graphicx}

\begin{document}
\def \nn{$N\overline{N}~~$}
\def \ee{$e^{+}e^{-}~~$}
\def \DD{$D^0\overline{D^0}~~$}
\def \D0bar{$\overline{D^0}~~$}
\def \cm{\,{\rm cm}}
\def \eV{{e\kern-.07em V}}
\def \GeV{\,{\rm G\eV}}
%%%%%%%%%%%%%%%%
%\begin{frontmatter}
% \begin{flushright}
%   FERMILAB-Pub-01/081-E 
% \end{flushright}
\title{
 A FIRST DESIGN OF THE PEP-N CALORIMETER
}
\author{P.~ PATTERI \\
on behalf of the calorimetry group
\thanks{co\,authors\,:~
R.~Baldini-Ferroli, M.~Cordelli, P.~LeviSandri, A.~Zallo ({\bf INFN Frascati,Italy});
V.~Bidoli, R.~Messi, L.~Paoluzi, E.~Santovetti ({\bf INFN and University of Rome ``Tor Vergata'',Italy })
}\\
\it  Laboratori Nazionali di Frascati dell'INFN -  via E.~Fermi 40, Frascati
I-00044  }

\maketitle

\begin{abstract} 
  A  preliminary design of the PEP-N electromagnetic calorimeter is given.
The spatial, energy and time resolutions achievable using a KLOE type
electromagnetic calorimeter are presented. 
\end{abstract}

\section{Physics requirements and performance goals}
\vskip 3 truemm
\begin{figure}[htb]
\centering
\includegraphics*[width=3.0in,height=3.0in]{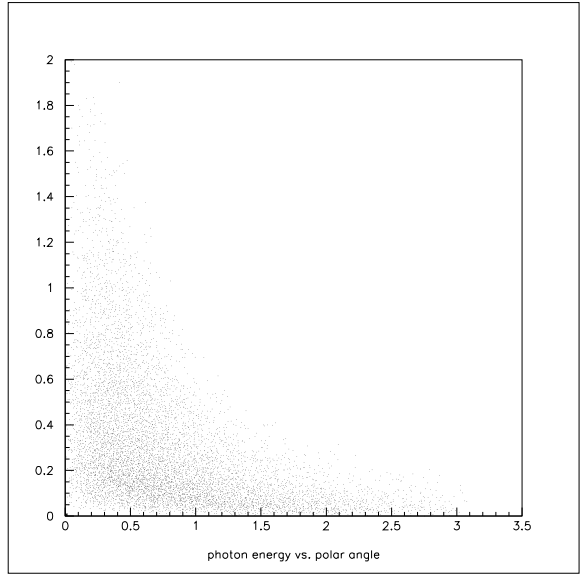}
 \caption{Angular (rad.) energy (GeV) photon correlation for 
\ee $\rightarrow 2\pi^+2\pi^-2\pi^0$.}
\label{fig:2dim}
\end{figure}
The main goals of the PEP-N physics programs are :
\begin{itemize}
\item
a measurement of R with a 2 $\%$ or better error.This condition requests
the measurement of all exclusive channels.
\item  
a measurement of the \nn  
form factors.
\item  
do spectroscopy in the energy range covered by the machine in order to
study  new  possible final states.
\end{itemize}
At PEP-N  energies, photons with energies lower that 100 MeV are produced,
 as shown in Fig. \ref{fig:2dim}
  for one of the
 channels with higher cross section 
 \ee $\rightarrow 2\pi^+2\pi^-2\pi^0$.
In order to achieve these  goals
 the electromagnetic calorimeter should be:
\begin{itemize}
\item   hermetic.
\item   have high efficiency, till 20 MeV, for charged and neutral particles.
\item have  good (few $\%$) energy resolution for photons.
 \item have good time resolution in order to separate \nn events from
other final states.
\end{itemize}
\par\noindent 
The KLOE electromagnetic calorimeter\cite{didomenico} was designed to detect  photons
in the (20-500) MeV energy range with good time resolution, so  
 its performances have been taken 
as guidelines in the design of the PEP-N electromagnetic calorimeter.
This detector can provide
 a fast and unbiased first level trigger,with high acceptance 
for final states with  low energy photons.A good K$/\pi$
separation is achievable and also some kind of $\pi/\mu$
discrimination is  possible.
\par
\subsection{The KLOE electromagnetic calorimeter}

\begin{figure}[htb]
\centering
\includegraphics*[width=3.5in,height=3.5in,angle=0]{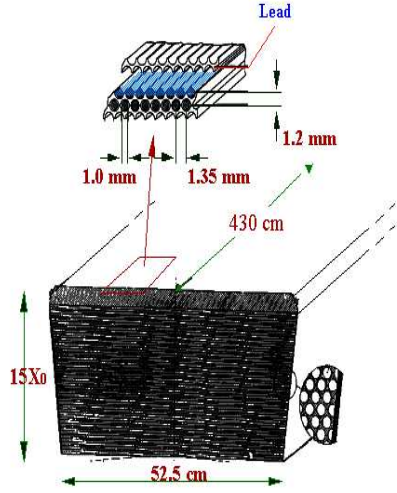}
 \caption{Sketch of a lead-scintillating fiber module.}
\label{fig:pbscifi}
\end{figure}

\noindent
The KLOE calorimeter is a fine sampling lead and
scintillating fibers calorimeter.   
The barrel mo\-du\-les  have trapezoidal cross
  section, 4.3 m long, 60 cm wide
  and 23 cm thick.
Each module is obtained gluing 0.5~mm thick lead foils worked to house
the 1~mm diameter fibers.  
The resulting structure (Fig. \ref{fig:pbscifi}) has
fiber:lead:glue volume ratio of 
42:48:10, an average density of $5~{\rm g/cm}^3$,
a mean radiation length  $1.5~{\rm cm}$,
and a sampling fraction of $\sim~15\%$ for
mi\-ni\-mum ionizing particles.
\noindent
The readout granularity is $\sim$ (4.4~$\times$~4.4)~cm$^2$, for a total
number of 4880 read-out channels. A precision in measuring the
photon conversion point in the transverse plane of $\simeq 1$ cm has been 
achieved.The
coordinate along the fiber is measured using the relation:
%\beeq
%z = v_f \cdot \frac{\Delta T}{2}
%\label{zcalo}
%\eneq
z = v$_f \cdot \frac{\Delta T}{2}$,
with $\Delta T\,$
the time difference 
at the two module ends and $v_f$ the effective light 
propagation speed in the fibers.
The measured effective light
propagation speed is $v_f = 17.2 \cm$/ns. 
%goo agreement  with the refractive index of the fiber core ($n = 1.6$)
%and the bounce angle of the light traveling in the fiber ($\theta \sim 21^o$).
\par\noindent
The z resolution is $\sigma_{\rm z} = \frac{1.24 \cm}{\sqrt{E(\GeV)}}$. 
\noindent
The obtained performances (Fig. \ref{fig:resolutions}),are summarized as follows:
\begin{figure}[htb]
\centering
\includegraphics*[width=3.0in,height=3.0in,angle=0]{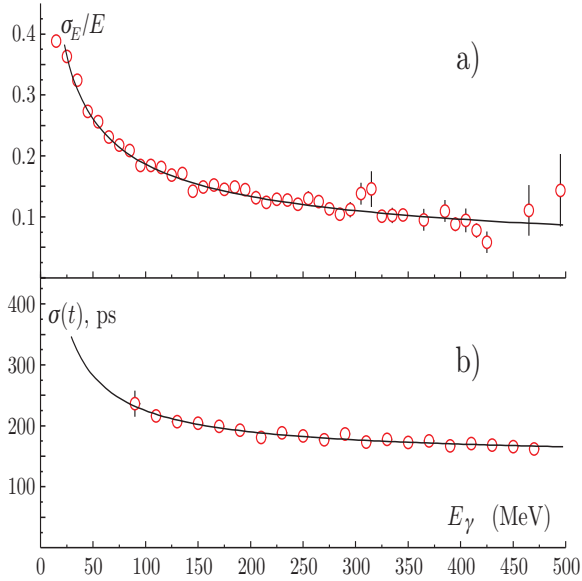}
 \caption{Energy and time resolutions as function of the photon energies.}
\label{fig:resolutions}
\end{figure}

\begin{itemize}
\item detection efficiency for photons with energy between 20 MeV and 500
  MeV of about 99$\%$.
\item resolution in the photon conversion point position in the transverse
  plane of $\simeq 1$ cm and in the z 
  coordinate  $\sigma_{\rm z} = \frac{1.24 \cm}{\sqrt{E(\GeV)}}$. 
\item energy resolution  $\frac{\sigma_E}{E}  \simeq \frac{5.7\%}{\sqrt{{\rm E(GeV)}}}$.
\item time resolution  $\sigma_t  \simeq \frac{54 \, {\rm ps}}{\sqrt{{\rm E(GeV)}}}+110{\rm ps}$.
\end{itemize}
\noindent
The calorimeter is inside a 0.6 T magnetic field with a consequent
residual magnetic field up to 0.2 T in the photomultipliers (PM) area and
with an angle with respect to 
PM axis up to 25$^{\circ}$; for this reason
fine mesh photomultipliers, Hamamatsu R5946, were specially designed for
KLOE.

\section{The PEP-N electromagnetic calorimeter}
\begin{figure}[htb]
\centering
\includegraphics*[width=3.0in,height=3.0in,angle=0]{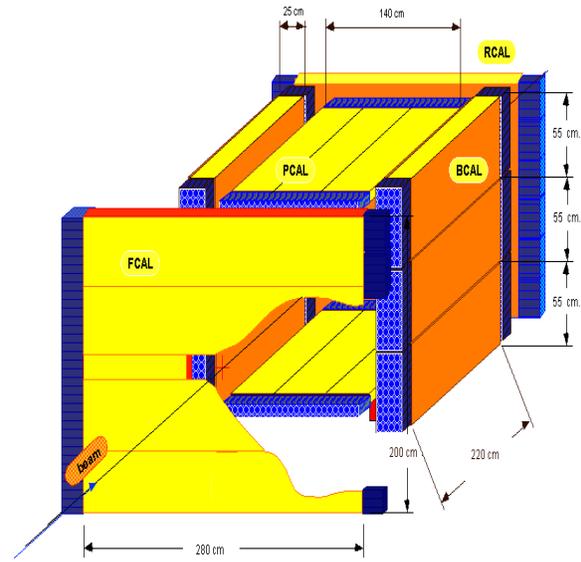}
 \caption{Exploded view of the PEP-N calorimeter.}
\label{fig:exploded}
\end{figure}

\begin{figure}[htb]
\centering
\includegraphics*[width=3.0in,height=3.0in,angle=0]{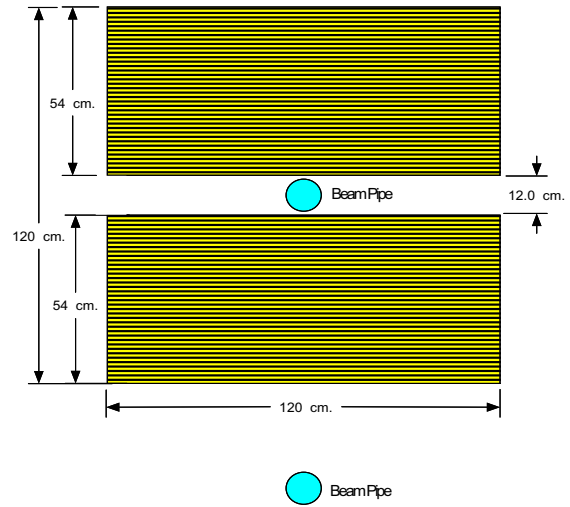}
 \caption{ RCAL detector.}
\label{fig:back}
\end{figure}

In Fig. \ref{fig:exploded} an exploded view of the calorimeter design
is shown.
It consists of barrel, forward and backward detectors.
\begin{itemize}
\item Each of the vertical sides of the  {\it barrel} have 3 mo\-du\-les  with rectangular cross
  section,  220 cm long, 55 cm high
  and 25 cm thick,  with  fibers parallel to the beams (BCAL detector). 
   Also the horizontal sides have 3 modules  220 cm long, 50 cm wide, 15cm thick, positioned
 over and under the TPC chamber, in order to complete the
coverage of the azimuthal acceptance (PCAL detector).
The angular region covered by the barrel modules is 
 27$^{\circ}~<~\theta~<~135^{\circ}$.
Due to lack of space PCAL detector is only 15 cm thick. The efficiency
and energy resolution, simulated with Montecarlo, are 
shown in Fig. \ref{fig:pcaleff} and
Fig. \ref{fig:pcalres}.
The efficiency is greater than 99$\%$ for energies higher then 40~MeV.
The gamma energy resolution  is 
 $\frac{\sigma_E}{E}  \simeq
  \frac{11\%}{\sqrt{{\rm E(GeV)}}}$ 
   (Fig. \ref{fig:pcalres}).
\begin{figure}[htb]
\centering
\includegraphics*[width=3.0in,height=3.0in,angle=0]{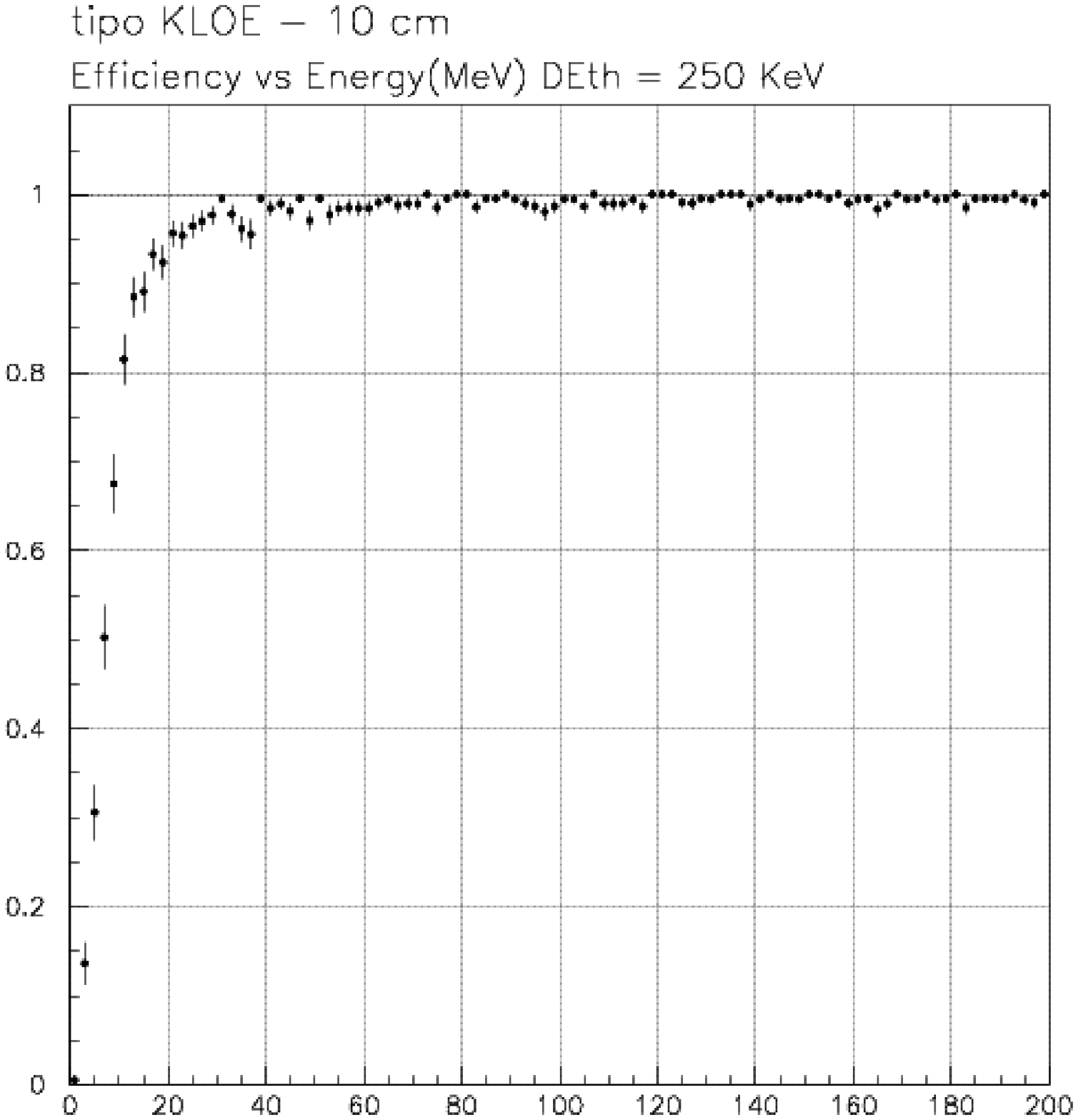}
 \caption{ Efficiency for PCAL detector.The 
$\gamma$ energies are in MeV.}
\label{fig:pcaleff}
\end{figure}
\begin{figure}[htb]
\centering
\includegraphics*[width=3.0in,height=3.0in,angle=0]{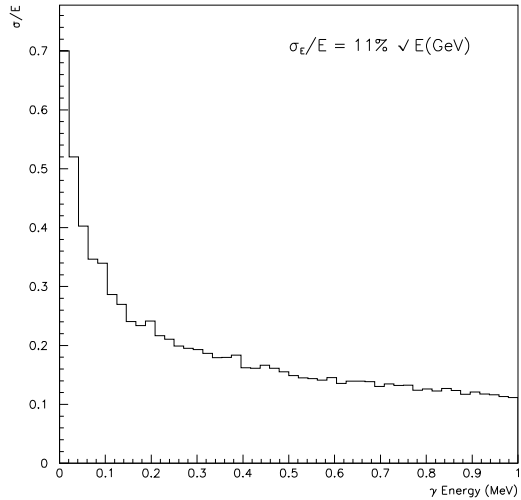}
 \caption{ Energy resolution for  PCAL detector.}
\label{fig:pcalres}
\end{figure}
\item The {\it forward detector }(FCAL )  is made of  mo\-du\-les (Table 1) with  the same thickness as BCAL. They are located at (130$~<~z~<~155$)~cm,  
cover an area of 280*180cm$^2~$, with polar angle range 
~6$^{\circ}~<~\theta~<~27.5^{\circ}$ (Fig. \ref{fig:forward}). 
\begin{figure}[htb]
\centering
\includegraphics*[width=3.in,angle=0]{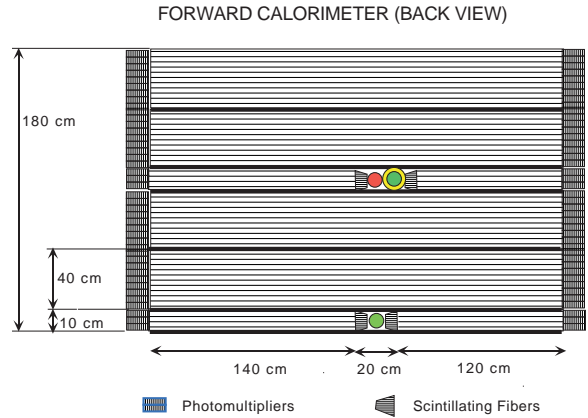}
 \caption{FCAL detector.}
\label{fig:forward}
\end{figure}
\item The {\it backward detector} (RCAL ) is composed of 2 modules positioned 
at z= -140 cm. They are 120 cm long, 54 cm high and 15 cm thick
 (Fig. \ref{fig:back}), specifically
 designed to detect very low energy photons.
The total area covered is ~120*120~cm$^2$  
 (135$^{\circ}~<~\theta~<~180^{\circ}$).
\end{itemize}
\par
\begin{figure}[htb]
\centering
\includegraphics*[width=3.0in,height=3.0in,angle=0]{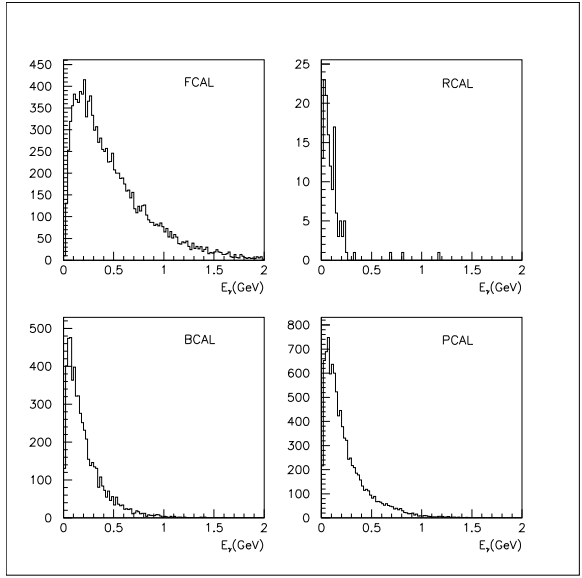}
 \caption{Photon energy distribution in calorimeter detectors.}
\label{fig:gammaenergy}
\end{figure}

 
In Table~\ref{tab:unfolded}  the dimensions, number of modules and number 
of photomultipliers requested are summarized. 
\begin{table}\caption{Sizes(cm),modules and photomultipliers for each detector.
\label{tab:unfolded}
}
\begin{center}
\begin{tabular}{|c|c|c|c|c|c|}  \hline \hline 
 & DX & DY & DZ & Modules &  Photom. \\ \hline
PCAL  & 50 & 15 & 220  & 6 & 60*6 \\ \hline           
BCAL & 25 & 55 & 220  & 4 & 110*4 \\ \hline                       
FCAL & 280 & 40 & 25  & 4 & 80*4 \\             
 & 120 & 10 & 25  & 1 & 20*1 \\             
 & 100 & 10 & 25  & 1 & 20*1 \\            
 & 140 & 10 & 25  & 2 & 20*2 \\ \hline            
RCAL & 120 & 54 &10  & 2 & 68*2 \\  \hline \hline           
\end{tabular}
\vfill
\end{center}
\end{table}
We have assumed the same readout granularity as KLOE,
 that is $\sim$ (4.4~$\times$~4.4)~cm$^2$, so that the total
number of  photomultipliers is 1350. With this configuration  a precision in measuring the
photon conversion point in the transverse plane of $\simeq 1$ cm should be achieved.
The inclusive photon  energy distributions 
in the four detectors, at E$_{\rm cms}$ = 2.25 GeV, is shown in Fig. \ref{fig:gammaenergy}.
\section{K-$\pi$ separation using PEP-N calorimeter}
\begin{figure}[htb]
\centering
\includegraphics*[width=3.in,height=4.in,angle=90.]{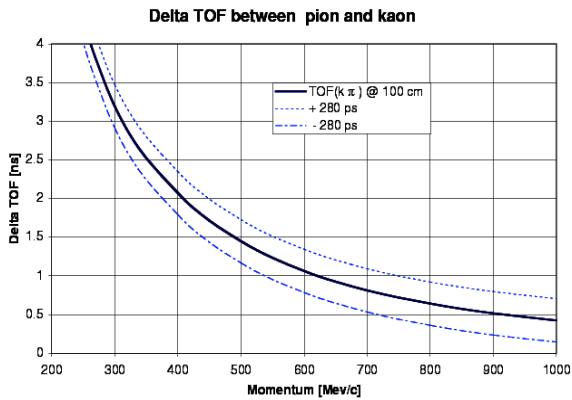}
 \caption{K-$\pi$ separation as function of  momentum 
after 1~m of path lenght.}
\label{fig:deltatov}
\end{figure}

\begin{figure}[t]
\centering
\includegraphics*[width=3.0in,height=3.0in,angle=0]{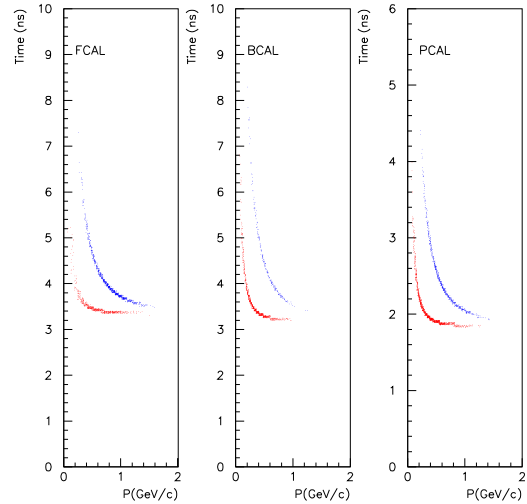}
 \caption{Time(ns) - momentum (GeV/c) correlation for KK$\pi\pi$ events for 
 FCAL,BCAL and PCAL at E$_{\rm cms}$ = 2.25 GeV.}
\label{fig:time}
\end{figure}

The PEP-N detector is equipped with an aerogel detector for K-$\pi$ 
separation in the (0.6$~<~P_{tot}~<~1.6$)GeV/c momentum range. 
The TPC chamber can be used to separate particles with momenta lower 
then 0.6 GeV/c (dE/dX measurement).Unfortunately 
the TPC measures badly the dE/dX of  particles hitting the pole calorimeter
PCAL where, due to lack of space, it is difficult to insert a specific 
particle identification
detector. Because of very good time resolution,better then 0.2~ns for m.i.p.,
PCAL can supply these informations. 
 Fig. \ref{fig:deltatov} shows the  K-$\pi$ separation
as function of the momentum after 1~m of path lenght.
Pions and kaons are well separated till 1~GeV/c momenta after 1~m of path.
In Fig. \ref{fig:time} the time-momentum separation for the process
KK$\pi\pi$ is shown 
for FCAL,BCAL and PCAL at  E$_{\rm cms}$ = 2.25 GeV.A 3 $\sigma$
separation is shown for these process till 0.8~GeV/c momenta.   

\section{ CONCLUSIONS}
\par
The present preliminary design of the electromagnetic calorimeter
fulfills all the physics requirements
for PEP-N.The energy resolution is acceptable, even compared to more expensive
crystal detectors.The excellent timing resolution is a very nice feature that
could be used to reduce background contributions and do also particle identification.
The very high efficiency for very low energy photons gives the possibility 
to build a minimum bias first level trigger.

\end{document}